\documentclass[aps,prl,showpacs,twocolumn,groupedaddress]{revtex4}
\begin{document}

\newcommand{\calA}{{\cal A}}
\newcommand{\calL}{{\cal L}}
\newcommand{\calM}{{\cal M}}
\newcommand{\calU}{{\cal U}}

\newcommand{\pU}{p_{\cal U}}
\newcommand{\dU}{d_{\cal U}}
\newcommand{\AdU}{A(d_{\cal U})}
\newcommand{\Abar}{\bar{A}(d_{\cal U})}

\newcommand{\avec}{{\bf a}}
\newcommand{\bvec}{{\bf b}}
\newcommand{\kvec}{{\bf k}}
\newcommand{\pvec}{{\bf p}}
\newcommand{\qvec}{{\bf q}}
\newcommand{\rvec}{{\bf r}}
\newcommand{\rvechat}{\hat{\bf r}}
\newcommand{\Svec}{{\bf S}}
\newcommand{\sigmavec}{{\bf \sigma}}
\newcommand{\vvec}{{\bf v}}

\newcommand{\kslash}{/\!\!\!\!\!k}
\newcommand{\pslash}{/\!\!\!\!\!p}
\newcommand{\qslash}{/\!\!\!\!\!q}

\newcommand{\oP}{\textrm{o-Ps}}
\newcommand{\pP}{\textrm{p-Ps}}
\newcommand{\tr}{\textrm{tr}}
\newcommand{\Br}{\textrm{Br}}

\newcommand{\ubar}{\bar{u}}
\newcommand{\vbar}{\bar{v}}
\newcommand{\sigmabar}{\bar{\sigma}}


\title{Long-range electron spin-spin interactions from
unparticle exchange}

\author{Yi Liao}
\email[]{liaoy@nankai.edu.cn}
\affiliation{Department of Physics,
Nankai University, Tianjin 300071, China}

\author{Ji-Yuan Liu}
\affiliation{Department of Physics, Nankai University, Tianjin
300071, China}

\date{Proofread version to appear in Phys. Rev. Lett.}

\begin{abstract}
Unparticles as suggested by Georgi are identities that are not
constrained by dispersion relations but are governed by their
scaling dimension, $d$. Their coupling to particles can result in
macroscopic interactions between matter, that are generally an
inverse nonintegral power of distance. This is totally different
from known macroscopic forces. We use the precisely measured
long-ranged spin-spin interaction of electrons to constrain
unparticle couplings to the electron. For $1<d<1.5$ the axial
vector unparticle coupling is excluded; and for $1<d<1.3$ the
pseudoscalar and vector couplings are also ruled out. These bounds
and the ones for other ranges of $d$ exceed or are complementary
to those obtained previously from exotic positronium decays.
\end{abstract}


\maketitle

Gravity and electromagnetism are the only known fundamental
interactions extending to a macroscopic distance. Due to its basic
importance, it has been a long tradition to search for extra
long-ranged interactions (for a recent work, see
\cite{Hoyle:2004cw}). Most experiments, especially those seeking for
deviations from the gravitational inverse square law, are only
sensitive to spin-independent interactions. A microscopic
spin-dependent interaction, which must be feeble to evade direct
detection in particle physics experiments if it exists at all, would
be simply averaged out for macroscopic bodies. To circumvent the
decoherence effects, one has to utilize spin-polarized samples
\cite{Moody:1984ba}. Although these are relatively new developments,
they already yield interesting and unique information beyond
spin-independent experiments (for experimental and theoretical
reviews, see e.g., \cite{ritter98,Dobrescu:2006au}, respectively).

In the language of quantum field theory (QFT), long-ranged
interactions are mediated by massless force quanta, the photon for
electromagnetism and the graviton for gravity. In the
nonrelativistic (NR) limit, the interaction potential always
starts in the form of $r^{-1}$, where $r$ is the separation of the
interacting particles. This is a joint result of the two facts in
QFT that all particles including force quanta are constrained by
dispersion relations quadratic in momentum and that we live in a
three space. When the spins of interacting particles enter or when
the small effects from relativistic corrections or simultaneous
multiple exchange of quanta are considered, higher integral powers
of $r^{-1}$ are also present.

It is common in theories beyond the standard model that there
exist hypothetical particles which have a mass tiny in the sense
that its Compton wave-length could be macroscopic. These particles
could then exert a force at a macroscopic distance. In the sense
of interactions, there is no surprise: they always follow an
inverse integral powers of distance up to an exponential factor.
These cover novel theories such as compactified extra dimensions
where the size of extra dimensions provides an effective mass in
four-dimensional theories.

So, what else can we imagine of a macroscopic force? The next
simplest or least strange would be a nonintegral power law. What
kind of force quantum could mediate such a force? It cannot be a
particle excitation, as we discussed above. This may partly
explain why such a force has not yet been analyzed. We must
confess at this point that we are so used to the concept of
particle that it is hard to move a step away from it.
Nevertheless, very recently Georgi has made an interesting
suggestion for an identity that is not a particle, dubbed
unparticle \cite{Georgi:2007ek}. He proposed a scenario showing
how such an identity could appear as a low-energy degree of
freedom from a scale invariant fundamental theory at high energy,
such as the one studied in \cite{Banks:1981nn}. The unparticles
must interact with ordinary matter, however feebly, to be
physically relevant. These interactions can be well described in
effective field theory (EFT) though it is generally difficult to
calculate from a fundamental theory.

An unparticle is an identity that does not enjoy mass as one of
its intrinsic properties. Namely, it is not constrained by a
dispersion relation as for a particle of mass $m$ and momentum
$p$, $p^2=m^2$. Instead, its kinematic property is defined by its
scaling dimension, $d$, which is generally a nonintegral number.
Scale invariance essentially determines its state density and via
unitarity its propagating property, up to a normalization factor
\cite{Georgi:2007ek,Georgi:2007si,Cheung:2007ue}. If the
normalization is fixed by analogy to the phase space of a system
of massless particles, the unparticle with a nonintegral $d$ could
be virtualized as $d$ number of invisible massless particles
\cite{Georgi:2007ek}.

The lack of a dispersion relation and the existence of a generally
nonintegral scaling dimension make unparticles sharply different
from particles. It is the purpose of this Letter to demonstrate
that an unparticle can mediate a long-range force between
particles of a nonintegral inverse power of distance depending on
$d$. We stress that this is an excluding characteristic of
unparticles that cannot be mimicked by any other theory of
particle physics, and is thus of fundamental importance. An
experimental indication of such a potential would definitely point
to unparticle physics and help discover a scale invariant
fundamental theory at high energy. Inversely, by employing
experimental constraints on extra macroscopic forces, this sets
bounds on the energy scale of unparticle physics. These bounds
could be more stringent than those from precision QED tests
\cite{Liao:2007bx}, because a feeble interaction between single
particles can be coherently amplified by a macroscopic mass if the
force is long-ranged.

Additional surprises have been unveiled previously. Due to lack of
a dispersion relation, a kinematically forbidden one-to-one
particle transition of different masses becomes possible for a one
particle to one unparticle transition \cite{Liao:2007bx}. For a
nonintegral $d$, the propagator gets a nontrivial phase in the
timelike region. This produces unusual interference patterns in
some processes \cite{Georgi:2007si}, and serves as a `strong
phase' to help discern CP violating effects \cite{Chen:2007vv}.
The studies so far have focused on unparticle effects at colliders
\cite{Georgi:2007si,Cheung:2007ue,Ding:2007bm,Mathews:2007hr,Chen:2007qr,
Ding:2007zw}, precision QED tests
\cite{Cheung:2007ue,Luo:2007bq,Liao:2007bx}, flavor-changing
neutral current processes
\cite{Luo:2007bq,Chen:2007vv,Aliev:2007qw,Li:2007by,Lu:2007mx,
Choudhury:2007js,Aliev:2007gr,Zhou:2007zq,Chen:2007je},
interactions with Higgs bosons \cite{Fox:2007sy}, in gauge boson
scattering \cite{Greiner:2007hr} and in astrophysics
\cite{Davoudiasl:2007jr}. Some theoretical issues are addressed in
Refs. \cite{Stephanov:2007ry,Fox:2007sy}.

We shall restrict ourselves to the system of electrons although we
are aware that there are constraints involving nucleons. The
reason is theoretical; for nucleons we have to study unparticle
interactions with quarks and gluons to make direct connection to
theory, which are then converted with unavoidable uncertainties to
interactions of nucleons. This implies in turn that we should
focus on the spin-dependent part of the electron interaction since
the spin-independent interaction of macroscopic samples is
dominated by that of nucleons. The leading interactions in EFT of
the electron ($\psi$) and unparticles are
\begin{eqnarray}
\calL_{\textrm{int}}&=&
C_S\overline{\psi}\psi\calU_S+C_P\overline{\psi}i\gamma_5\psi\calU_P
\nonumber\\
&&+C_V\overline{\psi}\gamma_{\mu}\psi\calU^{\mu}_V
+C_A\overline{\psi}\gamma_{\mu}\gamma_5\psi\calU^{\mu}_A,
\end{eqnarray}
which will induce long-ranged interactions between electrons. Here
$\calU_{S,P,V,A}$ stand for the fields of scalar, pseudoscalar,
vector and axial vector unparticles respectively. For simplicity,
we assign to them the same scaling dimension, $d$. The couplings
can be parametrized by $C_{S,P,V,A}=\pm
c_{S,P,V,A}\Lambda_{S,P,V,A}^{1-d}$, where $\Lambda_i,~c_i$ are
unknown energy scales and dimensionless positive numbers
respectively. One could set $\Lambda_i\sim 1$ TeV, say, and
constrain $c_i$, but we find it simpler to put $c_i=1$ and work
with $\Lambda_i$. The two can easily be converted into each other
indeed.

The propagator for a spin-0 unparticle is
\cite{Georgi:2007si,Cheung:2007ue}
\begin{eqnarray}
\frac{A_d}{2\sin(\pi d)}\frac{i}{(-p^2-i\epsilon)^{2-d}},\\
A_d=\frac{16\pi^{5/2}}{(2\pi)^{2d}}
\frac{\Gamma(d+\frac{1}{2})}{\Gamma(d-1)\Gamma(2d)},
\end{eqnarray}
with $p$ being the momentum. For a vector or an axial vector
unparticle, we attach a tensor projector for its spin. For the
vector one, it is immaterial whether to include the
$p_{\mu}p_{\nu}$ term since it vanishes due to current
conservation. For the axial one, however, there is no similar
conservation law. For definiteness, we shall simply work with
$-g_{\mu\nu}$. Note that theoretical considerations prefer a
narrow range for $d\in(1,2)$ \cite{Georgi:2007si}.

To obtain the potential between electrons, it is sufficient to
work out the $t$-channel electron scattering amplitude. We shall
keep terms up to $O(m^{-2})$ in the NR expansion where $m$ is the
electron mass, while higher order terms are suppressed at a
macroscopic distance. For this, we expand the kinetic term in
Schr\"odinger equation to the same relative order, as well as the
propagator and spinor bilinears \cite{lifshitz}. Ignoring terms
involving averaged velocities of the electrons in the center of
mass frame that are of no interest here, we obtain the potential:
\begin{eqnarray}
U^{--}_t(\rvec)=U^{--}_{\rm spin}(\rvec)+U^{--}_{\rm non}(\rvec),
\end{eqnarray}
where, extracting the common factors $A_dr^{1-2d}/(4\pi^2)$
\begin{equation}
\begin{array}{l}
U^{--}_{\rm spin}(\rvec)
=-C_A^2\Sigma_s\Gamma(2(d-1))\\
\displaystyle
+\frac{\Gamma(2d)}{4m^2r^2}\left[ (d-2)C_A^2\Sigma_s
-C_P^2\frac{\Sigma_s-(2d+1)\Sigma_a}{2(d-1)}\right.\\
\displaystyle
\left.+(C_A^2-C_V^2)\frac{(1-2d)\Sigma_s+(2d+1)\Sigma_a}{2(d-1)}
\right],\\
U^{--}_{\rm non}(\rvec)=(C_V^2-C_S^2)\Gamma(2(d-1))\\
\displaystyle
+\frac{\Gamma(2d)}{4m^2r^2}\left[(2-d)C_V^2-(3-d)C_S^2\right],
\end{array}
\end{equation}
and $\Sigma_s=\sigmavec_1\cdot\sigmavec_2,~
\Sigma_a=\sigmavec_1\cdot\rvechat\sigmavec_2\cdot\rvechat,~\rvechat=\rvec/r$
with the subscripts $1,2$ referring to the $e^-e^-$ pair. The
standard result for exchange of particles is recovered in the
limit $d\to 1$, up to contact terms proportional to
$\delta^3(\rvec)$. The latter cannot be obtained from the general
result because a simpleminded computation would give incorrectly
$\nabla^2r^{-1}\to 0$, although this is safe for $d\ne 1$.

Before we embark on the long-ranged interactions, we calculate the
hyperfine splitting (hfs) between the ortho- and parapositronium
ground state. There are two contributions to the $e^-e^+$
potential, one from the $t$-channel exchange, and the other from
the $s$-channel annihilation. The former, $U_t^{-+}$, is obtained
from $U_t^{--}$ by $C_V^2\to -C_V^2$. The latter gives in the NR
limit:
\begin{equation}
\begin{array}{l}
U_s^{-+}(\rvec)=A_d/[4\sin(\pi d)]
(-4m^2c^2-i\epsilon)^{d-2}\delta^3(\rvec)\\
\times\left[(3C_V^2+C^2_P+C_A^2)
+(C_V^2-C_P^2-C_A^2)\sigmavec_-\cdot\sigmavec_+\right],
\end{array}
\end{equation}
which is generally complex. Here the indices $\pm$ refer to the
$e^-e^+$ pair. Since the above is higher order than the
$t$-channel for $d<2$, we ignore it from now on.

Some comments are in order. Our main aim is to work out
long-ranged interactions of electrons. For this, the naive NR
expansion is suitable: higher terms will yield less important
terms. But for short-ranged bound state problems there is no
guarantee that higher terms make sense as they become more
singular than lower ones. This happens already in QED: the
expansion works well until terms of $(mc)^{-2}$ (with $c$ being
the velocity of light) because radiation enters only at
$O(c^{-3})$ \cite{lifshitz}. We will thus retain only the leading
term $\sim r^{1-2d}$ in $U_t^{-+}$. For $d\in(1,1.5)$, it behaves
well; for $d\in(1.5,2)$, it still yields a meaningful result for
the level shifts as long as it is treated as a perturbation,
although a negative potential singular than $r^{-2}$ results in
the phenomenon of falling-to-center. This is again similar to the
QED case.

After these considerations, the only term relevant for hfs is the
leading $C_A^2$ term. Using $\langle
r^{1-2d}\rangle=2^{2d-2}a^{1-2d}\Gamma[2(2-d)]$ for the
positronium ground state with $a=2/(m\alpha)$, we obtain the
relative shift:
\begin{equation}
\begin{array}{l}
E(1^3S_1)-E(1^1S_0)=-m\alpha^{2d-1}\displaystyle
\left(\frac{C_A}{m^{1-d}}\right)^2\frac{A_d}{2\pi^2}\\
\times\Gamma(2(d-1))\Gamma(2(2-d)),
\end{array}
\end{equation}
which is negative for $d\in(1,2)$. Note that the $s$-channel
contribution is lower by a factor $\alpha^{2(2-d)}$. The most
recent QED calculations \cite{Kniehl:2000cx,Melnikov:2000zz} yield
the value $+203.391~69(41,16)~{\rm GHz}$, to be compared with the
measured ones, $+203.387~5(16)~{\rm GHz}$ \cite{mills} and
$+203.389~10(74)~{\rm GHz}$ \cite{ritter84}. Since it is hard to
imagine higher order QED corrections can further reduce the
discrepancy, we suppose the gap is filled by the unparticle. Using
the most precise experimental value, we obtain
\begin{eqnarray}
\Lambda_A\ge 21~{\rm TeV}~\textrm{for}~d=1.5.
\end{eqnarray}
The bound decreases as $d$ increases.

Now we come to the macroscopic force mediated by unparticles. As
explained earlier, our main interest is in the spin-spin force
between electrons. To our knowledge, there are four precise yet
reliable experiments so far. Two of them used a torsion pendulum
\cite{ritter98,pan92}. They got a similar bound on anomalous
electron's spin-spin interaction that is less stringent by a
factor of 20 or 40 than those by experiments of induced
paramagnetization \cite{chui93,ni94}. In \cite{chui93}, a pair of
spin-polarized bodies made of ${\rm Dy}_6{\rm Fe}_{23}$ were used.
With all magnetic fields shielded and in the presence of an
anomalous spin-spin interaction, they would induce magnetization
in a paramagnetic salt sample made of ${\rm TbF}_3$. The anomalous
interaction is parametrized by a standard magnetic dipole-dipole
interaction with a global factor $\alpha_s$ measuring the relative
strength. They set the limit, $\alpha_s=(2.7\pm 2.4)\times
10^{-14}$ \cite{chui93}. In \cite{ni94}, another pair of
spin-polarized bodies made of ${\rm HoFe}_3$ were added and
aligned perpendicularly to the pair of ${\rm Dy}_6{\rm Fe}_{23}$.
There are now two kinds of signals as the table holding the bodies
rotates. The limit set from the new pair is, $\alpha_s=(-2.1\pm
3.5)\times 10^{-14}$. They combined the two to reach the final
limit:
\begin{eqnarray}
\alpha_s=(1.2\pm 2.0)\times 10^{-14}.
\end{eqnarray}

\begin{table}
\caption{Bounds on $\Lambda_{A,P,V}$ (in TeV) are shown as a
function of $d$. Data from Ref. \cite{ni94} are used with a
typical distance $r_0=25~{\rm cm}$. $\times$ stands for scales far
in excess of the Planck scale and $-$ for scales too low to be
useful.}
\begin{tabular}{|c|c|c|c||c|c|}
\hline $d$&$\log_{10}\Lambda_A$&$\log_{10}\Lambda_P$&
$\log_{10}\Lambda_V$&$d$&$\log_{10}\Lambda_A$\\
\hline
$1.2$&$\times$&$\times$&$\times$&$1.6$&$13.7$\\
$1.3$&$\times$&$6.44$&$5.77$&$1.7$&$9.04$\\
$1.4$&$\times$&$0.126$&$-0.307$&$1.8$&$5.53$\\
$1.5$&$20.3$&$-$&$-$&$1.9$&$2.81$\\
\hline
\end{tabular}
\end{table}

When employing the above limit, we should note the differences
between our interaction and the one used in fitting. Ours is
generally not of a standard dipole-dipole form in either the $r$
dependence ($r^{1-2d}$ or $r^{-1-2d}$ instead of $r^{-3}$) or the
relative weight of $\Sigma_{s,a}$ (not in a ratio of $1:(-3)$). An
accurate Monte-Carlo simulation based on our interaction is
certainly welcome, but this is not possible without detailed
knowledge of the samples and apparatus, especially their geometric
properties. Fortunately, due to the special arrangement in those
experiments, we can make reasonably good approximations. We note
that the magnetization direction of the salt lies in a plane
parallel to the plane of polarization of the spin-polarized
bodies. Their dimensions are much smaller than the vertical
separation between the salt and the bodies, and the bodies are
close to the apparatus' axis where the salt is placed. Considering
all of this, we expect that the spin-spin interaction between the
masses scales with the vertical distance up to an order one
geometric factor and that the $\Sigma_a$ term is much smaller than
$\Sigma_s$ because $\rvechat$ is very close to being perpendicular
to the spins for most pairs of the electrons in the salt and the
bodies. Isolating the $\Sigma_s$ terms whose coefficients are
constrained by $-0.8<\alpha_s\times 10^{14}<3.2$, we can set
bounds on $C_i$'s.

The largest contribution comes from the $C_A^2r^{1-2d}$ term with
others suppressed by a tiny factor of $(mr_0)^{-2}$, where $r_0$ is
the characteristic distance in the experiment. Since the term is
negative, we use the lower bound of $\alpha_s$ to get
\begin{equation}
\left(\frac{\Lambda_A}{\rm TeV}\right)^{2(d-1)}\!\!\!\ge 3.17
~10^4~\frac{\Gamma(d-\frac{1}{2})}{(2\pi)^{2d}\Gamma(d)}K^{2(d-2)},
\end{equation}
with $K=0.2~10^{-16}~{\rm cm}/r_0$. The bound is shown in table I
for a typical $r_0=25~{\rm cm}$. For $1.5<d<2$, $\Lambda_A$ is
very stringently bounded. [Equivalently, one could assume
$\Lambda_A\sim 1$ TeV and constrain $c_A$; for instance, at
$d=1.6$, one has $c_A<10^{-8}$.] For $1<d<1.5$, we have
practically $C_A\sim 0$ since $\Lambda_A$ is close to or exceeds
the Planck scale, so that we should consider the $O(m^{-2})$
terms. Though $C_{P,V}$ terms differ in sign and partly cancel, we
cannot gain more by treating them together because their scaling
dimensions are generally different. We choose to consider them one
by one and get separate bounds as follows:
\begin{eqnarray}
\left(\frac{\Lambda_P}{\rm TeV}\frac{1}{K}\right)^{2(d-1)}\ge %
6.07~10^{16}~
\frac{\Gamma(d+\frac{1}{2})}{(2\pi)^{2d}\Gamma(d)},\\
\left(\frac{\Lambda_V}{\rm TeV}\frac{1}{K}\right)^{2(d-1)}\ge %
1.52~10^{16}~
\frac{\Gamma(d+\frac{1}{2})(2d-1)}{(2\pi)^{2d}\Gamma(d)}.
\end{eqnarray}
The bounds are also shown in table I.

Unparticles result in a long-ranged force between matter, which is
generally an inverse nonintegral power of distance, most likely
between Coulomb and dipole ones. This is unique to unparticles and
cannot be disguised by particles in any other model conceived so
far. An experimental indication of it would unambiguously point to
unparticle physics and significantly modify our standard
conception of particle physics. On the other hand, existing
experiments on macroscopic electron's spin-spin interactions are
already useful in assessing the relevance of unparticles in our
world. The obtained pattern of constraints is complementary to
that in positronium decays \cite{Liao:2007bx}, and they together
constitute the best constraints worked out hitherto. This highly
restricts the relevance of unparticles in electron-involved
processes studied in the literature. For $1<d<1.5$, the axial
vector unparticle coupling is excluded. For $1.5\le d<2$, the
bound on it is much more stringent than in positronium decays. For
$1<d<1.3$, the pseudoscalar and vector unparticles couplings are
also ruled out. At $d\sim 1.5$ however, the bounds on them are
less stringent than from positronium decays. Since we are
restricted to spin-spin interactions, the scalar unparticle does
not set in at the considered order, which however is constrained
by positronium decays. Finally, we have studied the positronium
hfs due to unparticles. Although this is best measured in
positronium spectroscopy in absolute precision, it cannot compete
with its decays or macroscopic experiments.

\begin{acknowledgments}
YL would like to thank Prof. G.T. Gillies for kindly providing a
copy of Ref \cite{ritter98} which has been very helpful for his
understanding of the experimental status.
\end{acknowledgments}

\bibliography{}

\begin{thebibliography}{100}

\bibitem{Hoyle:2004cw}
  C.~D.~Hoyle {\it et al}., 
  Phys.\ Rev.\  D {\bf 70}, 042004 (2004).

\bibitem{Moody:1984ba}
  J.~E.~Moody and F.~Wilczek,
  Phys.\ Rev.\  D {\bf 30}, 130 (1984).

\bibitem{ritter98}
R.C. Ritter, G.T. Gillies, and L.I. Winkler, in {\it Spin in
Gravity}, edited by P.G. Bergmann, V. de Sabbata, G.T. Gillies,
and P.I. Pronin (World Scientific, Singapore, 1998), pp. 199-212.

\bibitem{Dobrescu:2006au}
  B.~A.~Dobrescu and I.~Mocioiu,
  J. High Energy Phys. 11 (2006) 005.

\bibitem{Georgi:2007ek}
  H.~Georgi,
  Phys.\ Rev.\ Lett.\  {\bf 98}, 221601 (2007).

\bibitem{Banks:1981nn}
  T.~Banks and A.~Zaks,
  Nucl.\ Phys.\  B {\bf 196}, 189 (1982).

\bibitem{Georgi:2007si}
  H.~Georgi,
  Phys.\ Lett.\  B {\bf 650}, 275 (2007).

\bibitem{Cheung:2007ue}
  K.~Cheung, W.~Y.~Keung, and T.~C.~Yuan,
  Phys.\ Rev.\ Lett.\ {\bf 99}, 051803 (2007).

\bibitem{Liao:2007bx}
  Y.~Liao,
  Phys.\ Rev.\  D {\bf 76}, 056006 (2007).

\bibitem{Chen:2007vv}
  C.~H.~Chen and C.~Q.~Geng,
  arXiv:0705.0689.

\bibitem{Ding:2007bm}
  G.~J.~Ding and M.~L.~Yan,
  Phys.\ Rev.\  D {\bf 76}, 075005 (2007).

\bibitem{Mathews:2007hr}
  P.~Mathews and V.~Ravindran,
  arXiv:0705.4599.

\bibitem{Chen:2007qr}
  S.~L.~Chen and X.~G.~He,
  arXiv:0705.3946 [Phys. Rev. D (to be published)].

\bibitem{Ding:2007zw}
  G.~J.~Ding and M.~L.~Yan,
  arXiv:0706.0325.

\bibitem{Luo:2007bq}
  M.~Luo and G.~Zhu,
  arXiv:0704.3532.

\bibitem{Aliev:2007qw}
  T.~M.~Aliev, A.~S.~Cornell and N.~Gaur,
  arXiv:0705.1326.

\bibitem{Li:2007by}
  X.~Q.~Li and Z.~T.~Wei,
  Phys.\ Lett.\  B {\bf 651}, 380 (2007).

\bibitem{Lu:2007mx}
  C.~D.~Lu, W.~Wang, and Y.~M.~Wang,
  Phys.\ Rev.\  D {\bf 76}, 077701 (2007).

\bibitem{Choudhury:2007js}
  D.~Choudhury, D.~K.~Ghosh, and Mamta,
  arXiv:0705.3637.

\bibitem{Aliev:2007gr}
  T.~M.~Aliev, A.~S.~Cornell, and N.~Gaur,
  J. High Energy Phys. 07 (2007) 072.

\bibitem{Zhou:2007zq}
  S.~Zhou,
  arXiv:0706.0302.

\bibitem{Chen:2007je}
  C.~H.~Chen and C.~Q.~Geng,
  Phys.\ Rev.\  D {\bf 76}, 036007 (2007).

\bibitem{Fox:2007sy}
  P.~J.~Fox, A.~Rajaraman, and Y.~Shirman,
  Phys.\ Rev.\  D {\bf 76}, 075004 (2007).

\bibitem{Greiner:2007hr}
  N.~Greiner,
  Phys.\ Lett.\  B {\bf 653}, 75 (2007).

\bibitem{Davoudiasl:2007jr}
  H.~Davoudiasl,
  Phys.\ Rev.\ Lett.\  {\bf 99}, 141301 (2007).

\bibitem{Stephanov:2007ry}
  M.~A.~Stephanov,
  Phys.\ Rev.\  D {\bf 76}, 035008 (2007).

\bibitem{lifshitz}V.B. Berestetskii, E.M. Lifshitz, L.P.
Pitaevskii, Quantum Electrodynamics (Butterworth-Heinemann,
London, 1996).

\bibitem{Kniehl:2000cx}
  B.~A.~Kniehl and A.~A.~Penin,
  Phys.\ Rev.\ Lett.\  {\bf 85}, 5094 (2000).

\bibitem{Melnikov:2000zz}
  K.~Melnikov and A.~Yelkhovsky,
  Phys.\ Rev.\ Lett.\  {\bf 86}, 1498 (2001).

\bibitem{mills}
A.P. Mills, Jr., and G.H. Bearman, Phys.\ Rev.\ Lett.\ {\bf 34},
246 (1975); A.P. Mills, Jr., Phys.\ Rev.\  A {\bf 27}, 262 (1983).

\bibitem{ritter84}
M.W. Ritter {\it et al}., Phys.\ Rev.\  A {\bf 30}, 1331 (1984).

\bibitem{pan92}
S.S. Pan, W.T. Ni, and S.C. Chen, Mod.\ Phys.\ Lett.\ A {\bf 7},
1287 (1992).

\bibitem{chui93}
T.C.P. Chui and W.-T. Ni, Phys.\ Rev.\ Lett.\ {\bf 71}, 3247
(1993).

\bibitem{ni94}
W.-T. Ni {\it et al}., Physica (Amsterdam) {\bf 194-196B}, 153
(1994).

\end{thebibliography}

\end{document}